# 2D materials coated plasmonic structures for SERS applications

Ming Xia[1]*

[1] Applied Materials Inc., Santa Clara, California 95054
* Correspondence: xiaming@g.ucla.edu

**Abstract:** Two-dimensional (2D) materials, such as graphene and hexagonal boron nitride, are new kind of materials that can serve as substrates for surface enhanced Raman spectroscopy (SERS). When combined with traditional metallic plasmonic structures, the hybrid 2D materials/metal SERS platform brings extra benefits, including higher SERS enhancement factors, oxidation protection of metal surface, and protection of molecules from photo-induced damage. This perspective gives an overview of recent progress in 2D materials coated plasmonic structure in SERS application. This review paper focuses on the fabrication of the hybrid 2D materials/metal SERS platform and their applications for Raman enhancement.

## 1. Introduction

Raman spectroscopy is an optical analysis technique providing characteristic spectral information of anlaytes and has a wide variety of applications in chemistry, biology and medicine [1] because of its capability of providing fingerprints of molecule vibration. One major drawback of Raman spectroscopy is the low yield of Raman scattering, leading to weak Raman signals in most cases, and thus normal Raman spectroscopy can hardly provide discernable signals of trace amount of analytes. Surface enhanced Raman spectroscopy (SERS) makes up this deficiency via plasmon resonance from metallic nanostructures. Molecules adsorbed at nanostructured metallic surface experience a large amplification of electromagnetic (EM) field due to local surface plasmon resonance, which leads to orders of magnitude increase in Raman yield and greatly enhanced Raman signal. SERS is capable of ultra-sensitive detection (single molecule detection) and allows for label-free detection with high degree of specificity[2-4]. To achieve high SERS enhancement factors, many efforts have been devoted to develop various metallic (mainly Au and Ag) nanostructures to enhance the local EM field [5-8].

Two-dimensional (2D) materials, such as graphene and hexagonal boron nitride (h-BN), have unique electronic and optical properties, and attract wide interests for potential applications in electronic devices, sensors, and energy generation[9-11]. In addition, 2D materials have been explored to enhance Raman signals[12-16]. Since the discovery of graphene's Raman enhancement capability[17], extensive researches have been done to reveal the enhancing mechanism of two-dimensional materials, as well as their application in Raman enhancement substrate[6, 8, 16, 18-20]. Unlike traditional SERS substrates, 2D materials provide a non-metallic surface to enhance Raman signal. Recently, combing 2D materials with metallic plasmonic structures to form a hybrid SERS platform becomes an emerging research field. 2D materials coated SERS platform offers synergetic Raman enhancement from both 2D materials and plasmon resonance, and other advantages such as metal oxidation protection and protection of molecules from photo-induced damages. This paper will first introduce the Raman enhancing mechanism of 2D materials, and then discuss the recent process of 2D materials coated plasmonic structures for SERS application, including their fabrication, sensitivity and stability.

## 2. Raman enhancement of 2D materials

This section will briefly introduce the Raman enhancement mechanism of 2D materials, including graphene, h-BN and molybdenum disulfide ($MoS_2$). Unlike EM enhancement mechanism of most metallic SERS substrates, Raman enhancement of 2D materials is due to chemical enhancement mechanism[12, 17, 21, 22]. Chemical enhancement factor on metallic surface is usually low (~10 to 100)[23] compared with EM enhancement factor (~$10^6$ to $10^{11}$)[2, 24, 25]. In a broad perspective, chemical enhancement can be considered as modification of the Raman polarizability tensor of molecule upon its adsorption, which in turn enhances or quenches Raman signals of vibrational modes[26, 27]. 2D materials provide a superior platform to study the chemical enhancement mechanism because they have no dangling bonds in vertical direction and have atomically flat surface.

Graphene is the first one explored to enhance Raman signals of molecules[17]. Raman enhancement of pristine graphene is ascribed to the ground state charge transfer mechanism[21]. In ground state charge transfer, analyte molecules do not form chemical bond with SERS substrate necessarily. Graphene is chemically inert and the charge transfer between molecules and graphene causes change in analytes' electronic distribution. Ground-state charge transfer can easily happen between graphene and molecules adsorbed on its surface because of graphene's two unique features: abundant π electrons on its surface and continuous energy band. Figure 1 (a) shows the proposed ground state charge-transfer process in graphene enhanced Raman system. In normal Raman scattered process, molecule absorbs photon energy and electrons are excited to a higher-energy level. The electrons then relaxes down the vibrational sub-structure and Raman scattered photons are emitted. The graphene electrons involvement in the Raman scattered process can enhance the electron−phonon coupling, and thus induce the enhancement of the Raman signals. The vibrational mode involving the lone pair or π electrons, which has stronger coupling with graphene, has highest Raman enhancement[15, 28].

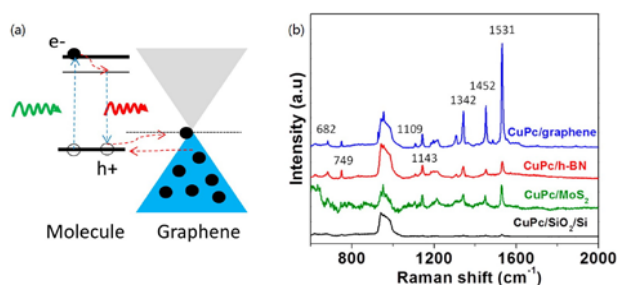

Figure 1. (a) Schematic of the Raman scattered process of graphene enhanced Raman spectroscopy[21]. (b) Raman spectra of the CuPc molecule on the blank $SiO_2$/Si substrate, on graphene, on h-BN, and on $MoS_2$ substrates. The numbers marked on the peaks are the peak frequencies of the Raman signals from the CuPc molecule[12].

h-BN and $MoS_2$ are other two kinds of 2D materials with different electronic and optical properties from graphene. h-BN is highly polarized and insulating with a large band gap of 5.9 eV [29]. CuPc molecule Raman signal is found to be enhanced by h-BN substrate. One proposed Raman enhancement mechanism of h-BN is the interface dipole interaction with analyte molecules, which causes symmetry-related perturbation in the CuPc molecule[12]. In addition, the Raman enhancement factor does not depend on the h-BN layer thickness, because the distribution of the intensity is uniform no matter how thick the h-BN flake is. Atomic layer thin $MoS_2$ is semiconductor and also has a polar bond[30]. For $MoS_2$, both the charge transfer and interface dipole interaction are much weaker compared with graphene and h-BN respectively. The Raman enhancement of $MoS_2$ is not as obvious as that of graphene and h-BN, as shown in Figure 1 (b).

**3. Two dimensional materials coated plasmonic nanostructures**

Traditional SERS analysis relies on metallic nanostructures that can generate strong local EM field. When combining 2D materials with metallic structure, the hybrid SERS substrate can provide even higher SERS enhancement factor due to the synergic effect of electromagnetic and chemical enhancement. 2D materials, like graphene and h-BN, could offer chemically inert and biocompatible surface[31, 32], which is favorable in bio-detection. With 2D materials as shielding layer on metallic surface, metal SERS platform such as Ag could be protected from oxidation and have longer shelf life, which can improve the stability and repeatability of SERS analysis. The following discussion will focus on the fabrication, sensitivity, and stability of 2D materials/plasmonic structure for SERS application.

**3.1 Fabrication**

2D materials/plasmonic structures require incorporation of 2D materials and metallic plasmonic structures that can provide high local electric field upon laser excitation. Common fabrication methods of

2D materials include mechanical exfoliation, chemical exfoliation, and chemical vapor deposition (CVD). Summary of 2D materials synthesis[33-35] and metallic SERS substrate[36] fabrication can be found elsewhere. This section will focus on the incorporation of 2D materials with metallic plasmonic structure.

One simple way to incorporate 2D materials with metallic nanostructure is to transfer CVD grown 2D materials on metal surface. Graphene and $MoS_2$ have been proven to be capable of overlapping on Au nanostructures and generating strong Raman signals of graphene and $MoS_2$ [37]. Zhu et al.[38] fabricated graphene-covered gold nanovoid arrays using CVD grown graphene and investigated the SERS performance of graphene/plasmonic structure. Figure 2 shows the graphene transfer process and the SEM images of graphene-covered gold nanovoid arrays. In this study graphene is actually suspended on Au nanovoid arrays instead of being conformally coated on Au surface. To achieve 2D material conformally coated SERS substrates, metallic structures need to have certain morphology. For instance, nanopyamid and nanocone structure can be conformally coated with 2D materials although some ripples are unavoidable. Figure 3 shows graphene coated Au nanopyramid [16] and $MoS_2$ coated $SiO_2$ nanocone[39], where 2D materials are transferred with the assistance of poly (methyl methacrylate) (PMMA). Metallic plasmonic structure with conformally coated 2D materials can be better isolated from air and thus has longer stability.

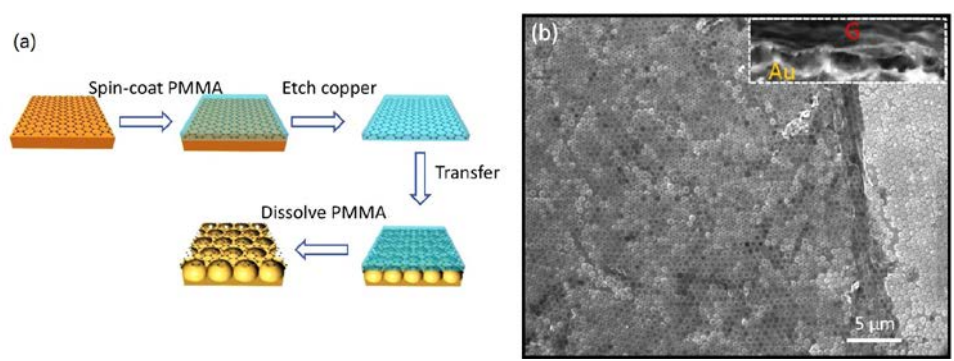

Figure 2. (a) Schematic illustrations of the graphene transfer process. (b) SEM image of a large-area nanovoid array integrated with the transferred monolayer graphene. The dark region is covered by graphene. The inset shows a SEM image of the cross-section of graphene-covered nanovoids[38].

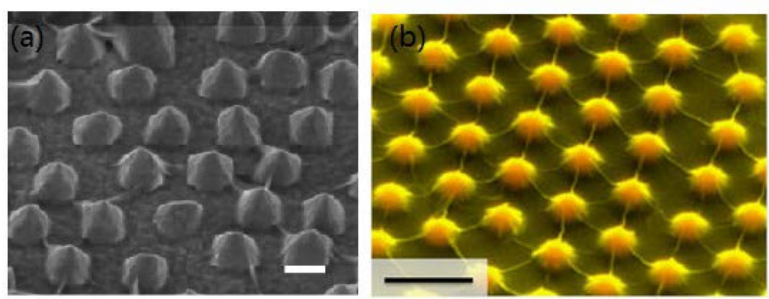

Figure 3. (a) Graphene coated Au nanopyramid structure. Scale bare is 200 nm [16]. (b) Tilted false-colour SEM image of the 2D strained $MoS_2$ crystal defined by the nanocone array. Scale bar is 500 nm [39].

PMMA assisted transfer method has advantage to coat 2D materials for plasmonic structures with various morphology. However, the drawback is that PMMA residue left on the surface of 2D materials [40] may generate noisy Raman peaks and prevent analyte molecules directly adsorbed on the surface of 2D materials. Therefore, special care needs to be paid to avoid large amount of PMMA residue. Another disadvantage of this transfer method is that the capillary force during the drying process of 2D materials may tear apart the 2D materials and expose the metallic surface. Xu et al[13] developed a novel flexible graphene/plasmonic structure with PMMA as a carrying substrate for SERS application. In this study, PMMA was used to support a flat graphene surface instead of a sacrificing transfer layer. Figure 4 shows the fabrication process of the flexible graphene SERS tape.

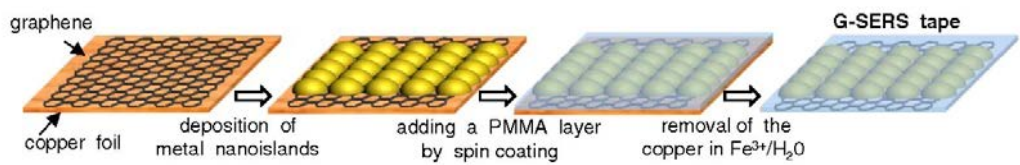

Figure 4. Schematic steps of the preparation route flexible G-SERS tape prepared from CVD-grown monolayer graphene [13].

Another way to incorporate 2D materials with metallic nanostructure is to use chemically exfoliated 2D materials to coat metallic nanoparticles. Kim at al.[41] developed a method to sandwich Ag nanoparticles between layers of reduced graphene oxide (rGO) and graphene oxide (GO) in order to prevent Ag nanoparticle from oxidation and boost Raman signals of analytes. Compared with CVD grown 2D materials, chemically exfoliated 2D materials are cost effective and easily to be functionalized [41]. Figure 5 shows the preparation of SERS substrates with chemically exfoliated graphene.

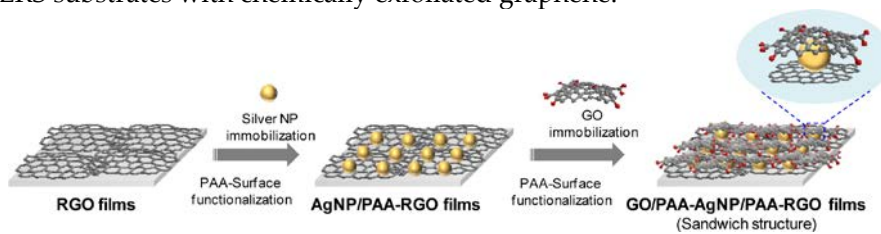

Figure 5. Fabrication Process of GO/PAA-AgNP/PAA-RGO Films for Application as SERS Platform[41].

Besides *ex-situ* transferring 2D materials on metal surface, *in-situ* grown 2D materials, like graphene and $MoS_2$, on metal surface is another attracting approach to incorporate 2D materials with plasmonic structure. Liu et al [42] developed a CVD process to grow graphene shell with controllable thickness on the surface of metal NPs. Figure 6 shows the fabrication process of graphene-encapsulated metal nanoparticles. *In situ* grown 2D materials on metal surface does not require 2D materials transfer process and has less chance to have polymer residue left on the surface of 2D materials. CVD *in-situ* grown 2D materials is a promising method to conformally coat 2D materials on metallic surface. However, due to the high temperature of CVD process, pre-designed metallic nanostructure may change their morphology during high temperature process and lose the pre-designed high local EM field. Low temperature plasma enhanced CVD method could be a potential choice to *in-situ* grow 2D materials on metal surface.

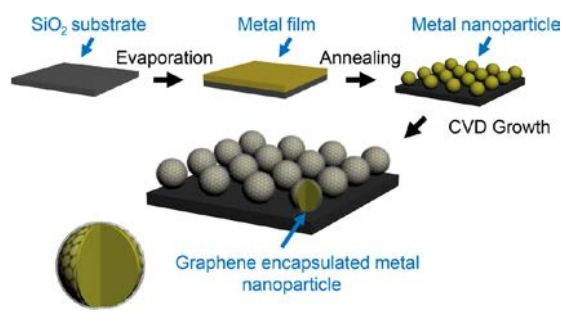

Figure 6. Production process for the Metal@Graphene to serve as a SERS-active substrate[42].

**3.2 Sensitivity**

Among various 2D materials, graphene is the most widely explored one to incorporate with plasmonic structure. Graphene/metal hybrid SERS platform shows superior SERS performance compared with bare metal SERS substrates. As a result of chemical interaction between graphene and target molecules, certain SERS modes are enhanced or prohibited. Although the chemical enhancement factor of 2D materials is not as high as metallic nanostructure, several tens' times of Raman signal enhancement could be essential when detecting molecules at single molecular level. Several times enhancement determines whether the Raman peaks can be seen or not. Wang et al[15] developed a graphene/Au nanopyramid hybrid SERS platform, which shows single-molecule detection capability for analytes like R6G and lysozyme. Even for molecules with small Raman cross-section, like dopamine and serotonin, graphene/Au hybrid platform can still achieve detection limit of $10^{-9}$ M in simulated body fluid[16]. With graphene/Au nanopyramid hybrid SERS substrates, serotonin molecule Raman peak hot spots and graphene peak hot spots actually coincide as seen from the Raman intensity mapping of analytes peak with that of the graphene G peak (Figure 7). The results indicate that the intrinsic Raman signal of 2D materials in 2D materials/metal hybrid SERS platform can serve as a gauge of the near-field EM-field intensity to locate hot spots. This unique feature of hybrid platform offers an advantage for molecule detection in ultra-low concentration. Actual hot spots of SERS substrates are rare and random even for patterned nanostructure. For extremely diluted solution, the spatial coincidence of molecules and hotspots is rare, leading to long time of up to hours spent on searching for measurable signals. With 2D materials' intrinsic Raman peak intensity as a SERS enhancement factor marker, the hot spots of the 2D materials/metal hybrid SERS platform could be located in advance and speed up the later detection of target molecules. For 2D materials used in hybrid SERS platform with patterned metallic SERS nanostructures, graphene is the ideal choice because graphene only has a few intrinsic Raman peaks and large-area high quality graphene is easily achievable. In addition, monolayer graphene has only 2.3% absorption of the incident laser, and its plasmon resonance frequency is the tetra Hz regime. Therefore, it has little effect on the EM field of underneath metallic SERS substrates.

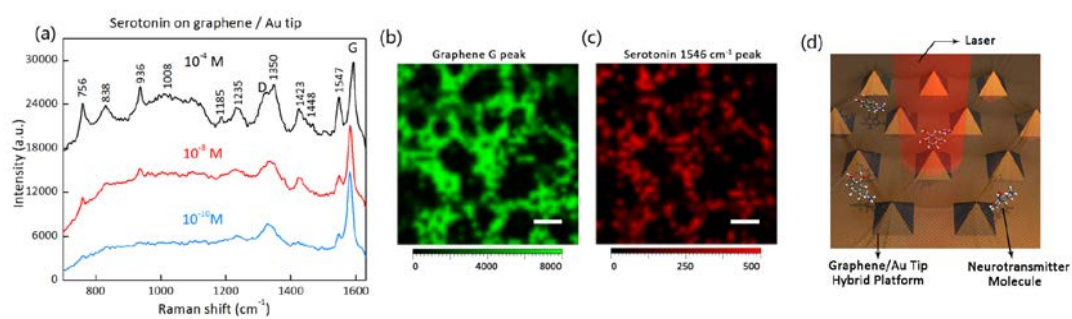

Figure 7. (a) Raman spectra of Serotonin molecules on graphene hybrid structure with 3 different concentrations ($10^{-4}$, $10^{-8}$, and $10^{-10}$ M). (b and C) Raman intensity mapping of graphene G band (green) and Raman intensity mapping of serotonin peak at 1546 cm$^{-1}$ (red) of the same area, scale bar, 10 μm. (d) schematic of graphene/Au nanopyramid SERS substrate [16].

Besides graphene, h-BN also served as coating layer on plasmonic structure for SERS application. Kim et al[43] reported h-BN layer wrapped Au nanoparticles as SERS substrate. h-BN coated Au SERS substrate can provide sensitive detection of aromatic hydrocarbon (PAC) molecules, such as B($\alpha$)P. PAC molecule Raman detection is very difficult using conventional metallic SERS because the weak interaction between polycyclic aromatic hydrocarbon (PAC) molecules and the metal surface prohibits their adsorption on the metal surface. With h-BN wrapped Au SERS substrates, noticeable and characteristic bands of B($\alpha$)P can be detected (Figure 8a), which is because the π−π interaction between B($\alpha$)P and h-BN enlarges the surface adsorption coverage (Figure 8b).

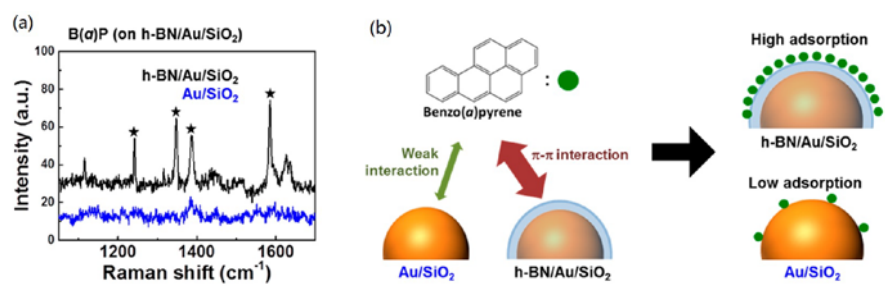

Figure 8. (a) SERS spectra benzo($\alpha$)pyrene on h-BN/Au/SiO$_2$ and Au/SiO$_2$ substrates. (b) Schematic mechanism to explain SERS of benzo($\alpha$)pyrene on h-BN/Au/SiO$_2$ and Au/SiO$_2$ substrates [43].

**3.4 Stability**

Ag nano-structure is known to have excellent SERS performance with wider plasmonic spectral window than other metallic structure made of Au or Al. However, one major weakness of Ag nanostructure is that it is easily to be oxidized in ambient environment. The degradation of Ag will lower the SERS performance and cause uncertainty of analysis. In addition, photo induced damage on analyte molecules is a well know side effect of metallic SERS substrates. This section will discuss recent process of using 2D materials as shielding layer to protect SERS metal substrates from oxidation and protect analyte molecules from photo-induced damages.

2D materials, like graphene and h-BN, are able to protect metal to be oxidized[44, 45]. This feature of 2D materials can also be used in SERS substrate development[42, 46]. When single layer graphene combines with Ag nanostructure, the hybrid SERS platform provides both better SERS performance and excellent stability in a harsh environment (sulfur) and at high temperatures (300 °C)[47]. Liu at al.[46] combined CVD grown graphene with silver SERS substrates and demonstrated that with the graphene as protecting layer, the hybrid graphene/Ag SERS substrate could achieve large-area uniformity and long-term stability. Li et al.[48] compared the oxidation protection effect between CVD grown graphene and rGO coated Ag nanoparticles. They found out that CVD-grown monolayer graphene served as a better protecting layer than rGO to effectively suppress the oxidation of Ag nanoparticles. As seen from Figure 9, CVD grown graphene coated Ag SERS substrate can provide stable R6G SERS signals up to 28 days with ambient aerobic exposure, while rapidly decreasing Raman signals are seen from rGO coated and bare Ag nanoparticles. Worse performance of rGO protected Ag nanoparticle is because 1) the wide size distribution of rGO results an incoherent thin film; and 2) the cracks and holes on rGO film could act as a channel to allow air reach Ag surface, leading to the oxidation of Ag nanoparticles.

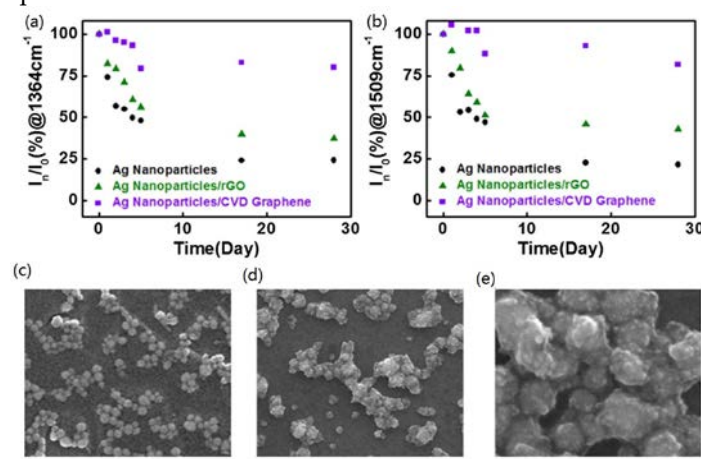

Figure 9. (a) The normalized intensity of the R6G Raman peak at 1364 cm$^{-1}$ collected using respectively the unprotected (black), rGO-protected (green) and CVD graphene-protected (purple) Ag nanoparticles as substrates, versus the time of aerobic exposure; (b) The normalized intensity of the R6G Raman peak at 1509 cm$^{-1}$ collected using respectively the unprotected (black), rGO-protected (green) and CVD graphene-protected (purple) Ag nanoparticles as substrates, versus the time of aerobic exposure; (c) SEM image of the CVD graphene-protected Ag nanoparticles after their 28-day use as the SERS substrate for the measurement of R6G; (d and e) SEM image of the unprotected Ag nanoparticles after their 28-day use as the SERS substrate for the measurement of R6G [48].

Another benefit to combine 2D materials with metallic nanostructures is that 2D materials can help protect molecules from photo-induced damage, such as photobleaching[13, 42, 49, 50]. The photobleaching (or photodegradation) of the Raman anlaytes induced by the laser is a well-known side effect in SERS experiments, especially for dye molecules. When combining graphene with metallic nanostructure, the hybrid SERS platform is more stable against photo-induced damage with an even higher enhancement factor. Liu at al.[42] fabricated graphene-encapsulated metal nanoparticles for molecule detection, and found out that AuNP/graphene hybrid substrate could significantly suppress photobleaching and fluorescence of cobalt phthalocyanine (CoPc) and R6G molecules. For instance, within the 160 s measurement period, the 1534 cm$^{-1}$ peak intensity of CoPc molecules decreases dramatically for Au NPs, while the same peak intensity almost keep constant for Au@Graphene, as shown in Figure 10 (a) and (b). Zhao et al.[50] also demonstrated that graphene can enhance the photostability of R6G molecules with graphene coated Ag SERS substrates during continuous light illumination. Enhanced photostability of molecules provided by graphene during SERS detection is attributed to π−π interactions between graphene surface and molecules[42,50]. Molecule π−π interaction with graphene allows the charge transfer between graphene and molecules, providing additional path for the molecules to relax from the excitation state to the ground state[51]. This process reduces the number of molecules at excitation states and thus decreases photobleaching rate. Similar protection effect can be achieved by using h-BN layer as well. Kim et al.[43] reported a h-BN film wrapped Au substrate showing extraordinary stability against photothermal and oxidative damages during laser excitation, as shown in Figure10 (c) and (d). This outstanding stability against photothermal damage of h-BN wrapped Au SERS substrate is attributed to the ultrafast heat dissipation through the h-BN layer. With 2D materials as a shielding layer, hybrid SERS substrates will provide long-term stability.

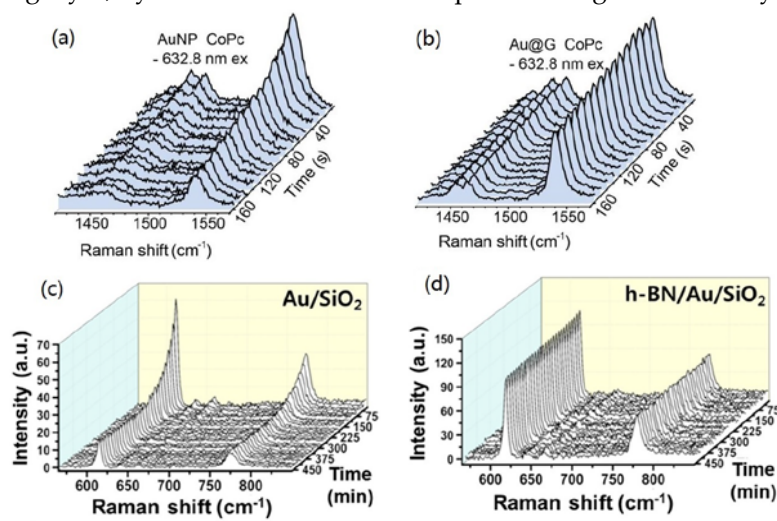

Figure 10. Stability of SERS signals of monolayer CoPc LB films on (a) Au and (b) Au@G[42]. Photothermal and chemical stability of 3L h-BN/Au/SiO2 substrate. SERS spectra of R6G on the Au/SiO2 substrate (c) without and (d) with h-BN protection at different time points (laser power = 0.1 mW, time interval = 15 min)[43].

## 4. Conclusion and perspective

In summary, 2D materials' Raman enhancement is due to chemical enhancement, which differentiates them from metallic SERS substrates. Coating 2D materials on metallic SERS substrates introduces extra benefits over bare metal substrates. First, adding 2D materials can further increase SERS enhancement factor due to the synergic effect of electromagnetic and chemical enhancement. Second, the atomic thin film of 2D materials can help map out the hot spots of metallic nanostructure without affecting the local EM field of metallic nanostructure underneath. For example, a Raman mapping of graphene G peak over the hybrid SERS substrates could give the precise position of hot spots. Finally, adding 2D materials as a shielding layer offers chemically inert surface and helps to reduce the fluctuation of SERS signal caused by degradation of the metallic nano-structures, photobleaching or metal-catalyzed site reactions, and thus improve the long-term stability and repeatability of SERS analysis.